\documentclass[12pt]{article}
\usepackage[latin2]{inputenc}
\usepackage{epsfig}
\usepackage{ifthen}
\begin{document}
\sloppy
\begin{flushleft}
DESY 98-076 \hfill ISSN 0418-9833\\
June 1998
\end{flushleft}
\vspace*{1.8cm}
\begin{center}
\begin{Large}
\boldmath
{\bf Di-jet Event Rates in Deep-Inelastic Scattering at HERA \\}
\end{Large}
\vspace*{1cm}
\begin{Large}
H1 Collaboration \\
\vspace*{0.5cm}
\end{Large}
 
\end{center}
\newcommand{\xbj}{\ensuremath{x_{\mathit{Bj}}}}
\newcommand{\rz}{\ensuremath{{\rm R}_2}}
\newboolean{figatend}
\setboolean{figatend}{false}
\begin{abstract}
\noindent Di-jet event rates have been measured for deep-inelastic 
scattering in the kinematic domain 5~\raisebox{-.8ex}
{$\stackrel{\textstyle<}{\sim}$} $Q^2$ \raisebox{-.8ex}
{$\stackrel{\textstyle<}{\sim}$} $100 \mbox{ GeV}^2 $ and $ 10^{-4}$
\raisebox{-.8ex} {$\stackrel{\textstyle<}{\sim}$} \xbj\ 
\raisebox{-.8ex} {$\stackrel{\textstyle<}{\sim}$} $10^{-2} $, and for jet
transverse momenta squared $p_t^2$ \raisebox{-.8ex} 
{$\stackrel{\textstyle>}{\sim}$} $Q^2$. The analysis is based on data 
collected with the H1 detector at HERA in 1994 corresponding to an integrated 
luminosity of about 2 pb$^{-1}$.
Jets are defined
using a cone algorithm in the photon-proton centre of mass system
requiring jet transverse momenta of at least 5 GeV. The di-jet event rates are
shown as a function of $Q^2$ and \xbj.
Leading order models of point-like interacting photons fail to describe the 
data. Models which add resolved interacting photons or which implement 
the colour dipole model give a good description of the di-jet event rate.
This is also the case for next-to-leading order calculations
including contributions from direct and resolved photons.

\end{abstract}

\vfill
\begin{center}
  Submitted to the European Physical Journal
\end{center}
\thispagestyle{empty} 

\newpage
\noindent
 C.~Adloff$^{35}$,                
 S.~Aid$^{13}$,                   
 M.~Anderson$^{23}$,              
 V.~Andreev$^{26}$,               
 B.~Andrieu$^{29}$,               
 V.~Arkadov$^{36}$,               
 C.~Arndt$^{11}$,                 
 I.~Ayyaz$^{30}$,                 
 A.~Babaev$^{25}$,                
 J.~B\"ahr$^{36}$
% ...

\vspace*{\baselineskip}

\noindent
 $ ^1$ I. Physikalisches Institut der RWTH, Aachen, Germany$^a$ \\
% ...
$ ^8$ Institut f\"ur Physik, Universit\"at Dortmund, Dortmund,
       Germany$^a$\\
% ...
$ ^{11}$ DESY, Hamburg, Germany$^a$ \\
% ...
$ ^{13}$ II. Institut f\"ur Experimentalphysik, Universit\"at Hamburg,
          Hamburg, Germany$^a$  \\
% ...
$ ^{23}$ Physics Department, University of Manchester, Manchester, UK$^b$ \\
% ...
$ ^{25}$ Institute for Theoretical and Experimental Physics,
          Moscow, Russia \\
 $ ^{26}$ Lebedev Physical Institute, Moscow, Russia$^{f,k}$ \\
% ...
$ ^{29}$ LPNHE, Ecole Polytechnique, IN2P3-CNRS, Palaiseau, France \\
 $ ^{30}$ LPNHE, Universit\'{e}s Paris VI and VII, IN2P3-CNRS,
          Paris, France \\
% ...
$ ^{35}$ Fachbereich Physik, Bergische Universit\"at Gesamthochschule
          Wuppertal, Wuppertal, Germany$^a$ \\
 $ ^{36}$ DESY, Institut f\"ur Hochenergiephysik, Zeuthen, Germany$^a$ \\
% ...
$ ^a$ Supported by the Bundesministerium f\"ur Bildung, Wissenschaft,
        Forschung und Technologie, FRG,
        under contract numbers 7AC17P, 7AC47P, 7DO55P, 7HH17I, 7HH27P,
        7HD17P, 7HD27P, 7KI17I, 6MP17I and 7WT87P \\
 $ ^b$ Supported by the UK Particle Physics and Astronomy Research
       Council, and formerly by the UK Science and Engineering Research
       Council \\
% ...
$ ^f$ Supported by the Deutsche Forschungsgemeinschaft \\
% ...
$ ^k$ Supported by Russian Foundation for Basic Researches 
       grant no. 96-02-00019 \\

\newpage
 
\section{Introduction}

The study of jets in deep-inelastic lepton-proton scattering (DIS)
provides a testing ground for perturbative QCD\@.  Partons emerging from
the scattering process manifest themselves as jets of collimated hadrons 
which are observable in the experiment.
 
In the naive
quark parton model, the virtual photon is absorbed by a single quark
(antiquark) of the proton resulting in one jet from the struck quark
and one from the proton remnant. Both jets have no transverse
momentum in the photon-proton centre of mass frame (cms), when neglecting 
the intrinsic motion of the partons inside the proton. To first order in
$\alpha_s$, the leading order (LO) for di-jet production, two jets
with balanced transverse momenta in the photon-proton cms are
produced in the hard scattering process, in addition to the proton remnant
jet. The hard scattering can either be the quark initiated QCD-Compton (QCDC)  
or the gluon initiated photon-gluon fusion (BGF) process.
%
 
In this analysis we present a measurement of the fraction \rz\ 
of di-jet events in all DIS events, referred to as the di-jet rate.
It is presented as a function of
the Bjorken scaling variable \xbj, integrated over the virtuality of the
exchanged photon $Q^2$, and of $Q^2$, integrated over \xbj. Jets are
defined using a cone algorithm in the photon-proton cms requiring jet
transverse momenta of at least 5~GeV.  The measured jet rates are
corrected for detector effects.
 
Previous measurements of jet rates at HERA \cite{herajet} used the
JADE jet algorithm at photon virtualities $Q^2$ large compared to the
squared jet transverse momenta $p_t^2$. The present analysis probes a
region of jet phase space characterized by jet transverse momenta
squared of similar size or larger than the photon virtuality,
$p_t^2/Q^2$ \raisebox{-.8ex} {$\stackrel{\textstyle>}{\sim}$} 1. 
It has significantly better precision and has its emphasis on higher $Q^2$
compared to a previous study of single inclusive jet production \cite{tania}.
There the data were found to be in good agreement with LO QCD models which 
included a resolved partonic structure of the virtual photon that evolves with
$Q^2$.

In this study we investigate whether the di-jet rate can be described by 
LO QCD models with just point-like (direct) interactions of the virtual photon
and with models with additional contributions from resolved photons, which may 
be considered as an effective description of higher order QCD effects. 
We also consider the colour dipole model. Finally
our measurements are compared to next-to-leading order (NLO) QCD calculations 
which include either only direct or direct and resolved
virtual photons.
%
 
The measurement was performed using data taken in 1994 with the
H1 detector at the HERA storage ring, where 27.5 GeV positrons were
collided with 820 GeV protons.


 
\section{The H1 Detector} 

A detailed description of the H1 apparatus is given elsewhere
\cite{bib_h1}. The parts of the detector which are essential for this
measurement are the liquid argon (LAr) calorimeter \cite{lar1}, the
backward lead-scintillator calorimeter (BEMC) \cite{bemc}, and the
tracking chamber system.
 
\newcommand{\degree}{\ensuremath{^\circ}} The energy of the scattered
positron is measured in the BEMC which covers the range in polar angle%
\footnote{The polar angle $\theta$ is defined with respect to the
  positive $z$-axis, the proton beam direction.},
$\theta$, from 151\degree\ to 176\degree. It consists of stacks of lead and
scintillator plates with a total of 21.7 radiation lengths.  The BEMC
is laterally segmented into square modules of $16\times 16\,{\rm
  cm}^2$, with smaller modules at the inner and outer radii.  The
scintillation light is read out with photodiodes via wave length
shifters along two opposite sides of each module.  The absolute energy
scale was determined to a precision of 1\% \cite{h1-f2}. The energy
resolution is given by $\sigma_E/E=39\%/E \oplus 10\%/\sqrt{E} \oplus
1.7\%$ ($E$ in GeV) \cite{bemc}.

A cluster energy deposition exceeding a threshold of $\approx 7$\,GeV in the
BEMC was the primary trigger condition for events used in this
analysis.
 
The position of the scattered positron is measured with the backward
proportional chamber (BPC) located in front of the BEMC covering
the angular range
$155\degree < \theta < 174.5\degree$. The BPC consists of four layers
of wires strung vertically, horizontally, and at $\pm 45\degree$.  The
position resolution is $\sigma_{x,y}=1.5\,{\rm mm}$.
 
Hadronic energy is detected in the highly segmented ($\approx
45000$ channels) LAr calorimeter which extends from $4\degree < \theta
< 154\degree$. The depth of the LAr calorimeter varies between 4.5 and
8 hadronic interaction lengths in the region $4\degree < \theta <
128\degree$.  The uncertainty of the energy scale for hadrons
is 4\%. The hadronic energy resolution is
$\sigma_E/E=50\%/\sqrt{E} \oplus 2\%$ ($E$ in GeV), as measured with
test beams \cite{lar2}.
 
Charged tracks in the central region ($25\degree<\theta<155\degree$)
are measured with the central drift chamber system. Two jet chambers
with wires in the $z$-direction allow measurements of track positions
in the $r$-$\phi$-plane to a precision of $\sigma_{r\phi}=170\,{\rm
  \mu m}$. The $z$ coordinate is measured to a precision of
$\sigma_{z}=320\,{\rm \mu m}$ using drift chambers with
wires forming approximate circles around the beam. The momentum resolution is
$\sigma_{p_t}/p_t^2 < 1\%~{\rm GeV}^{-1}$.
 
The forward tracking detector covers $7\degree < \theta < 25\degree$
and consists of drift chambers with alternating planes of parallel
wires and others with wires in the radial direction. It allows the
measurement of track segments to a precision of $\sigma_{x,y}\leq 200\,{\rm
  \mu m}$. 

Two electromagnetic calorimeters located downstream in the positron beam
direction measure positrons and photons from the bremsstrahlung process
$ep \rightarrow ep\gamma$ for the purpose of luminosity determination.


\section{Data Selection}

The data sample used for the present analysis corresponds to an
integrated luminosity of 1.97 pb$^{-1}$ taken during the 1994 run
period.
The phase space region of DIS events considered in this analysis is
defined as follows:
\begin{eqnarray}
  156^\circ <& \theta_{e} & <  173^\circ       \nonumber \\
             &            E'& >  11{\rm\,GeV}   \label{psdis} \\
             &           y  & >  0.05       \nonumber
\end{eqnarray}
Here, $\theta_{e}$ is the polar angle of the scattered positron and
$E'$ is its energy.  The variables $\xbj$, $Q^2$ and $y$ (the
inelasticity variable)
are all determined from the 4-vector of the scattered positron.
This selection ensures that the
scattered positron is  well inside the acceptance region of the BEMC, that the
trigger efficiency is high, that the kinematic variables are well
reconstructed, and that photoproduction background and radiative
corrections are small. Photoproduction events, where the scattered
positron is not detected in the backward direction, form a background
if a particle from the hadronic final state entering the BEMC
is misidentified as the scattered positron.

Additional cuts are applied for the identification of the scattered
positron and to further suppress the influence of QED radiation and
photoproduction background \cite{h1-f2,JS}:
\begin{itemize}
\item The event must have a reconstructed vertex with a $z$ position
  within $\pm30$\,cm of the nominal position.
\item The candidate positron shower is required to have a small
  lateral spread by applying the cut $r_{\rm clust} < 5$\,cm, where
  $r_{\rm clust}$ is the energy-weighted mean transverse distance from
  the shower centre of gravity of each energy deposition sampled by the
  photodiodes.
\item There must be a BPC signal within 5\,cm of the straight line
  connecting the shower centre with the event vertex.
\item The quantity $\sum_i (E_i-p_{z,i})$, where the sum is over all
  calorimeter
  energy depositions in the final state, is expected to be equal to
  twice the positron beam energy.  An undetected positron in a
  photoproduction event or initial state photon radiation will
  decrease the value of this observable. For this analysis, $35 < \sum_i
  (E_i-p_{z,i}) < 70$\,GeV is required.
\end{itemize}
 
\section{Jet Reconstruction and Selection}

Jets are reconstructed using clusters of energy \cite{bib_h1} measured
in the LAr calorimeter. 
Cluster energies are corrected for the
difference in response to hadronic and electromagnetic energy deposition
and for losses due to dead
material and cracks. 
The cluster energy
and the direction from the
interaction point to the cluster centre are used to construct a
massless four-vector.

The calorimetric energy measurement can be improved for low energy particles
by using in addition to the energy the measured momentum of each charged
particle track. 
To avoid double counting of energy each track was allowed to
contribute at most 300\,MeV. The value of 300\,MeV was found to be
optimal for reconstructing the transverse momenta of jets in simulated
events \cite{JS}.

Jets are defined in this analysis using a cone algorithm \cite{cone}.
A cone is defined by a circular area of radius $R$ in the $\eta^*$ -- $\phi^*$ plane, where
$\eta^*$ and $\phi^*$ are the pseudo-rapidity%
\footnote{The pseudo-rapidity $\eta^*$ is given by $-\ln\tan(\theta^*/2)$.}
and azimuthal angle in the photon-proton cms%
\footnote{Variables measured in the photon-proton cms are marked by a
  ${}^*$. The positive $z$ direction is defined to be along the
  incident proton direction.}.
A jet candidate consists of all objects (clusters and tracks)
whose massless four vectors fall inside a cone. 
The jet transverse
momentum $p^*_{t,\rm jet}$ is calculated as the scalar sum of the transverse
momenta $p^*_{t}$ of the jet objects. 
The jet $\eta^*$ and $\phi^*$  (jet direction) are
calculated as the $p^*_{t}$-weighted averages of the $\eta^*$ and
$\phi^*$ of the objects. 
This way of calculating the jet parameters is usually called
the ``$p_t$''-scheme \cite{pt-scheme}.
An iterative procedure is used to find the jets of an event.
Initially, every object in turn is used to define the cone centre of a
candidate jet.
The jet directions of the candidate jets%
\footnote{Several initial cone centres
may result in the same candidate jet.}
are then used as
the cone centres for the next iteration.
This is repeated until the 
resulting jet directions are identical                   
to the cone centres.                                     
Then, also the midpoint in the $\eta^*$ -- $\phi^*$ plane of each pair
of jets is considered as a candidate jet centre, and the procedure
is repeated.
Jets which have more than a fraction $f$ of their $p^*_{t,\rm jet}$
contained in a higher transverse momentum jet are discarded. 
Finally, $p^*_{t,\rm jet}$ is required to exceed a minimum value
$p^*_{t,\rm min}$.

In this analysis, the following parameters were chosen:
\begin{equation}
  R=1, \quad p_{t,\rm min}^{*}=5{\rm\,GeV}, \quad \mbox{and } f =
  0.75. \label{psjets}
\end{equation}
Exactly two jets per event fulfilling these criteria are
demanded. In addition, the pseudo-rapidity difference $\Delta\eta^*$
of the two jets is required to be in the range
\begin{equation}
|\Delta\eta^*|<2. \label{psdeta}
\end{equation}
In leading order this cut is equivalent to requiring
$|\cos \hat{\theta}|<0.76$, where $\hat{\theta}$ is the polar angle
between the emerging and incoming partons in the parton-parton or
gamma-parton cms.
It separates the jets from the proton remnant.
The resolution in jet transverse momentum $\Delta p_t^*/p_t^*$ is
approximately 20\% at $p_t^* \sim 5$~GeV.
 
The number of di-jet events found is 4\,957 while the total number of
DIS events selected amounts to 112\,806.  To obtain the di-jet rate
$\rz$, the number of di-jet events is divided by the total number of
events in the same region of $\xbj$ and $Q^2$. $\rz$ is measured in
bins of $\xbj$, integrated over $Q^2$, and in bins of $Q^2$,
integrated over $\xbj$.


\section{Data Correction}

The residual background from photoproduction processes was determined
using the PHOJET Monte Carlo (MC) generator \cite{phojet} and was
separately subtracted from the total number of events and the di-jet events
as a function of $Q^2$ and \xbj. This generator has been proven to
give a good description of photoproduction background \cite{h1-f2}.
The correction for this background as well as the other corrections described 
below were obtained using MC events which were processed by the 
H1 detector simulation, reconstruction, and analysis chain.
The largest subtraction of the photoproduction background occurs in the 
lowest \xbj\
and $Q^2$ bins, where it amounts to 14\% and 9\% of the total event
sample respectively, and to 3\% and 1\% of the di-jet sample.  It is
below 5\% in the total sample and negligible in the di-jet sample in
all other bins.  For the di-jet rate, the correction is only significant
in the lowest \xbj\ and $Q^2$ bins where it increases the rate by
$\approx 10\%$.
 
Radiation of photons from the incoming or outgoing positron
leads to values of \xbj, $Q^2$, and $y$, as determined from the
scattered positron, which differ from the true kinematics of the
photon-proton interaction.%
\footnote{%
In the outgoing positron case, this only applies if the 
angle between the radiated photon and the outgoing positron is large.}
These effects are different for the total and the di-jet sample. They
were corrected using the DJANGO MC generator
\cite{django}. The correction factor on the di-jet rate was found to be 
1.08, independent of \xbj\ and $Q^2$.

\ifthenelse{\boolean{figatend}}{}{\begin{figure}
\begin{center}
\mbox{\epsfig{file=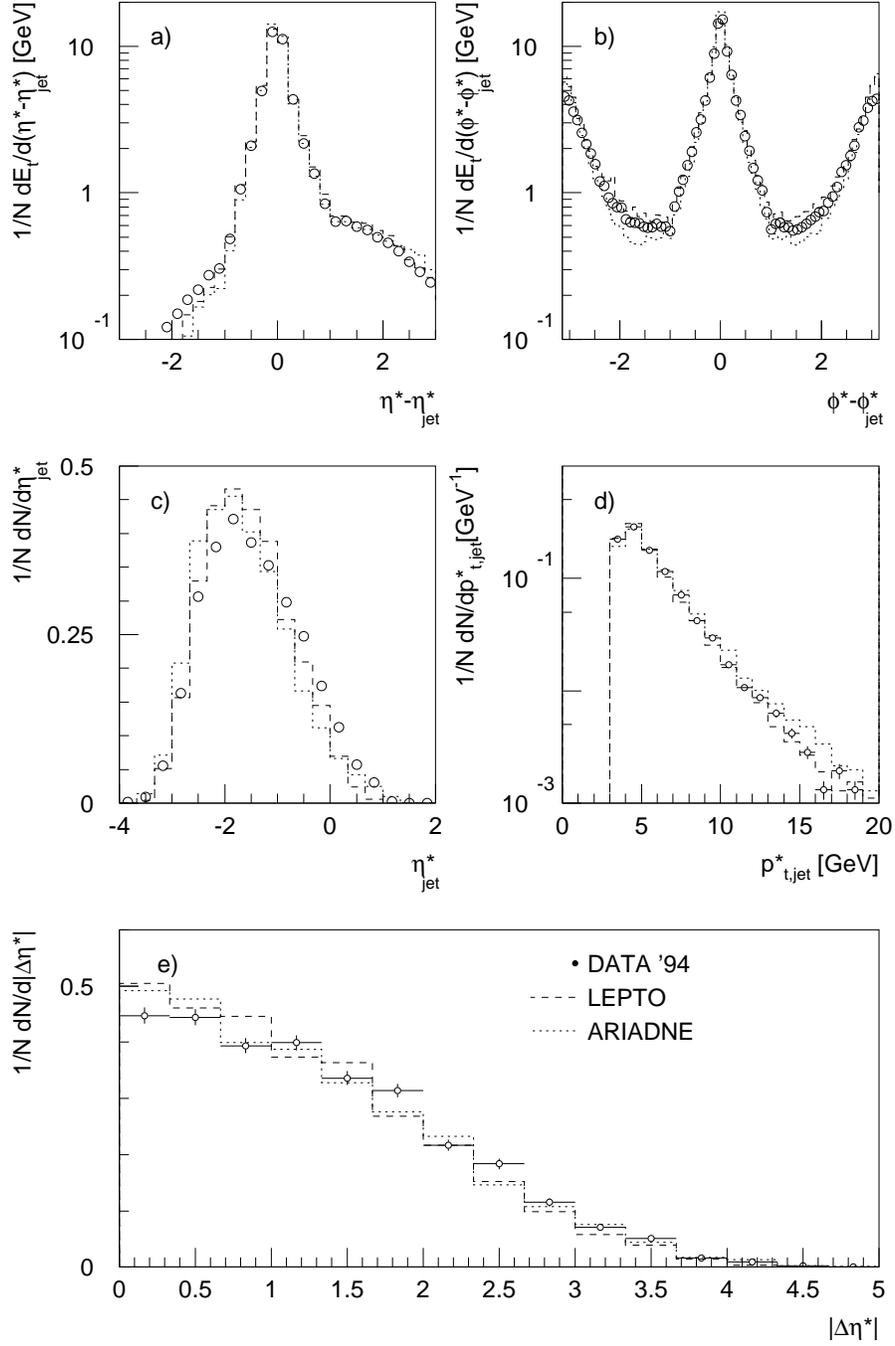,height=18cm}}
\end{center}
\caption{\label{figctl} {\it Transverse energy flow with respect to the
  jet axis, a) versus $\eta^*-\eta^*_{\rm jet}$ in a slice defined by
  $|\phi^*-\phi^*_{\rm jet}|<1$, b) versus $\phi^*-\phi^*_{\rm jet}$
  in a slice defined by $|\eta^*-\eta^*_{\rm jet}|<1$. In the Figs.~c) and d)
  the $\eta^*$ and $p_t^*$ spectra of the jets are shown. In Fig.~e)
  the $|\Delta\eta^*|$ distribution is displayed. Points denote data,
  the histograms indicate the distributions obtained from LEPTO
  (dashed) and ARIADNE (dotted). The curves are normalized to the
  number N of entries; there are two entries per di-jet event in
  Figs.~a)--d) and one in Fig.~e). For Fig.~d), $p_{t,\rm min}^{*}$ 
  (Eqn.~\ref{psjets})
  was lowered to 3.5\,GeV and for Fig.~e), the cut on $|\Delta\eta^*|$
  (Eqn.~\ref{psdeta}) was omitted.} }
\end{figure}

} 
The correction of the di-jet rate for detector acceptance and
efficiencies was performed with two MC models, LEPTO \cite{lepto64} 
and ARIADNE \cite{ariadne}, which will be discussed in section 7. They were
used with two sets of parton density parameterizations, MRS-H \cite{mrsh}
and GRV-94 HO \cite{grv} as implemented in PDFLIB \cite{pdflib}. The
average prediction of these
models was used to obtain bin-wise correction factors
$c=\rz^{\rm MC,hadrons}/\rz^{\rm MC,recon.}$ for the di-jet
rate $\rz$. 
This procedure is justified since the observables of the
jet events, which are sensitive to detector effects, are well
described by the Monte Carlo simulations.  Critical observables in this
sense are the energy flow within and around the jets, the $\eta^*$ and
$p_t^*$ distribution of the two jets and the pseudo-rapidity
difference $|\Delta\eta^*|$ between them.

Fig.~\ref{figctl}
shows a comparison of the experimental distributions with the two MC
models. Only the curves obtained with the \mbox{MRS-H} parton density
are shown, the curves with \mbox{GRV-94 HO} are very similar.  There
is good agreement between data and simulated events except for the
$\eta^*$ distribution of the jets.  On top of the bin-wise correction,
which ignores this small discrepancy, an additional correction for
this effect was applied \cite{JS}. 
It takes into account the fact that on average the $\eta^*$ of
the jets for data is higher than for MC which leads to an overestimation of 
the correction factors $c$, as the jet reconstruction efficiency depends
on $\eta^*$.  This correction reduces \rz\ by 10\% in the lowest and
by 1\% in the uppermost bins of $Q^2$ and \xbj.  The combined
correction factors vary between 1.0 for low \xbj\ and $Q^2$ values and 1.2
for high values.


\section{Systematic Errors}

Several sources of systematic uncertainties were investigated.  A
change in the hadronic energy scale of the LAr calorimeter by its
estimated precision of $\pm 4\%$ results in a global change of the
di-jet rates by ${}^{+9\%}_{-7\%}$.  The correction for radiative
effects has a global uncertainty of $\pm 3\%$ based on Monte Carlo
statistics.  These two errors were added in quadrature to give an
overall systematic error of ${}^{+10\%}_{-8\%}$.
 
Changing the energy scale of the positron measurement in the BEMC
within its $\pm 1\%$ uncertainty results in a change of \rz\ by
$\pm2\%$ in all $Q^2$ bins. In the \xbj\ bins, the change varies
between $\pm 1\%$ for the lowest and $\pm 9\%$ for the highest bin.
The systematic errors on the corrections for acceptance and efficiency
were obtained by using the maximal variation of the correction factor
for any particular model compared to the mean in each bin. They are of
the order of $10\%$. 
The additional corrections for the difference in the mean values of
$\eta^*$ between experiment and simulation have a systematic error of
the order of 2\%.
These errors were added in quadrature to give
a bin by bin systematic error. It varies between 5\% and 19\%.


\section{QCD Calculation of Di-jet Rates}

\ifthenelse{\boolean{figatend}}{}{\ifthenelse{\boolean{figatend}}{\begin{figure}[h]}{\begin{figure}}
\begin{center}
\mbox{\epsfig{file=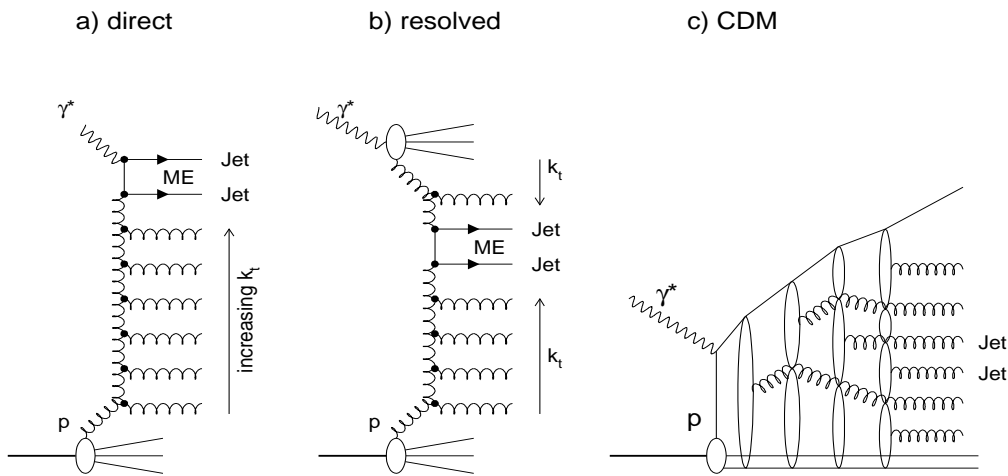,width=14cm}}
\end{center}
\caption{\label{figlad}
  {\it Generic diagrams of initial state parton emission in $ep$ scattering (a,b).
  In the direct process (a) the hardest emission
  given by the QCD matrix element occurs at the top of the ladder.
  Emissions down the ladder are ordered with decreasing transverse momenta $k_t$.
  In the resolved process (b) the hardest emission
  given by the QCD matrix element may occur anywhere
  in the ladder with increasingly soft emissions along the ladder
  towards the proton and the photon. In the colour dipole model (c) gluon 
  emissions are not ordered in transverse momentum $k_t$.} }
\end{figure}

}
Scattering processes involving the production of high $p_T$ partons
(hard scattering processes) 
are expected to be well described by perturbative
QCD\@. In this analysis hard collisions are selected by requiring two
jets with transverse momentum above 5 GeV.  We consider three different
QCD inspired
models labeled DIR (direct), DIR+RES (direct + resolved), CDM
(colour dipole model), and NLO QCD calculations
at the parton level for comparison with experimental data.
We briefly describe their most important features.
\begin{description}
\item[DIR:] The LO QCD matrix elements (BGF and QCDC) 
  are convoluted with the parton densities in the proton.  Only
  direct interactions of the photon are considered as indicated in
  Fig.~\ref{figlad}a. For this model we have used the LEPTO \cite{lepto64} and
  RAPGAP \cite{rapgap} MC programs. The latter has been
  used without generating diffractive (rapidity gap) processes.
  RAPGAP and LEPTO give results consistent with each other to better than
  10\%.
 
\item[DIR+RES:] In addition to the direct contribution discussed
  above, a contribution from quarks and gluons inside the photon is
  considered, as shown in Fig.~\ref{figlad}b.  This resolved photon
  contribution is assumed to set in
  only for scales of the hard subprocess \raisebox{-.8ex}
  {$\stackrel{\textstyle>}{\sim}$}$Q^2$.
  For the virtual photon parton densities the SaS-1D parameterization
  \cite{sasgam} was used. For this set of parton densities, the scale
  of the onset of the anomalous contribution in the virtual photon
  $P^2_0 = \mbox{max}(Q^2_0, Q^2)$ was chosen, where 
  $Q^2_0 = 0.36~\mbox{GeV}^2$ is the starting point of the $Q^2$
  evolution\footnote{this corresponds to the SaSgam parameter IP2=2}.
  The LO resolved photon contribution is implemented in the RAPGAP program.
  The contribution from longitudinal virtual photons is neglected.
\end{description}

In both the DIR and DIR+RES models we have used for the renormalization 
and factorization scale $\mu^2 = Q^2 + p_t^2$
as it provides a smooth transition between the DIS and photoproduction
regimes. Additional emissions in the initial
and final state are generated by parton showers \cite{parton_shower}
in the leading log DGLAP \cite{dglap} approximation. In this
approximation the radiated partons in the initial state are strongly
ordered in transverse momentum $k_t$\footnote{$k_t$ is the transverse
momentum relative to the proton and photon axis in the
photon-proton cms.}, with the hardest emission in the ladder
occurring next to the hard matrix element (Figs.~\ref{figlad}a and 
\ref{figlad}b). 
\begin{description}
\item[CDM:] In the colour dipole model \cite{cdm}, as implemented in
  the MC generator ARIADNE \cite{ariadne}, gluon emission originates
  from a colour dipole stretched between the scattered quark and the
  proton remnant.  Each emission of a gluon leads to two dipoles which
  may radiate further, generating a cascade of independently radiating
  dipoles (Fig.~\ref{figlad}c).  These gluons are not ordered in $k_t$. 
  A similar feature is found in the BFKL \cite{non-kt-bfkl} evolution scheme.

  The colour charge of the proton remnant (a di-quark in the simplest
  case) is assumed not to be point-like, leading to a phenomenological
  suppression of gluon radiation \cite{ariadne} in the direction of
  the remnant. This suppression occurs for hard gluons with
  wavelengths smaller than the size of the remnant. In addition, the
  colour charge of the scattered quark is taken to be extended,
  depending on the virtuality $Q^2$ of the photon (photon size
  suppression). This in turn leads
  to a suppression of radiation in the direction of the scattered
  quark \cite{ariadne,cdm_extended_quark}.

  The QCDC component of the di-jet rate
  depends in the CDM model on the size of the colour charge  while for 
  the DIR model it depends on the parton densities of the proton. 
  At low \xbj\ this results in a considerably
  enhanced di-jet rate for CDM compared to DIR \cite{rathsman}.  The
  photon-gluon fusion process, which is not naturally described by the
  CDM, is treated similarly as in the DIR approach discussed above.
\end{description}

\noindent In the DIR, DIR+RES, and CDM models, hadronization was
performed with the Lund string fragmentation scheme as implemented in
JETSET \cite{jetset}.
\begin{description} 
\item[NLO:] Finally, we consider two calculations in next to leading
  order (NLO) in the strong coupling constant $\alpha_s$ as implemented
  in the Monte Carlo integration programs DISENT \cite{disent} and JETVIP
  \cite{jetvip}. These programs provide cross sections for partons
  rather than a full hadronic final state. DISENT takes the soft
  and collinear divergencies arising in any NLO QCD calculation into
  account by using
  the subtraction method while JETVIP relies on the phase space slicing
  method. Both DISENT and JETVIP calculate NLO cross sections assuming
  a direct interacting photon. In addition, JETVIP
  provides a consistent calculation in NLO of direct {\it and}\/
  resolved interacting photons using parameterizations of the virtual
  photon structure functions.  For the latter, the SaS-1D parameterization 
  \cite{sasgam} (transformed to $\overline{MS}$) was used. For both
  the DISENT and JETVIP calculations 
  the factorization and renormalization scales were chosen to be 
  $\mu^2 = Q^2 + 50$~GeV$^2$, where 50~GeV$^2$ represents a good
  estimate of the 
  average transverse momentum squared of the jets in the hadronic 
  cms for the selection described before.
\end{description}
For comparing the corrected di-jet rate with models and parton level
calculations, we have used the CTEQ4M parameterization \cite{cteq} of parton
densities inside the proton with the corresponding 
$\Lambda^{(5)}_{\overline{MS}}$ of 202~MeV  
(different parton density parameterizations were used in the models used to
correct for detector effects, c.f.\ section 5).
The DIR, RES, and CDM models implement the one loop expression for the calculation
of $\alpha_s$ and in the DISENT and JETVIP programs the two loop expression 
was used.

%


\section{Results and Discussion}

\ifthenelse{\boolean{figatend}}{}{\begin{figure}
\begin{center}
\mbox{\epsfig{file=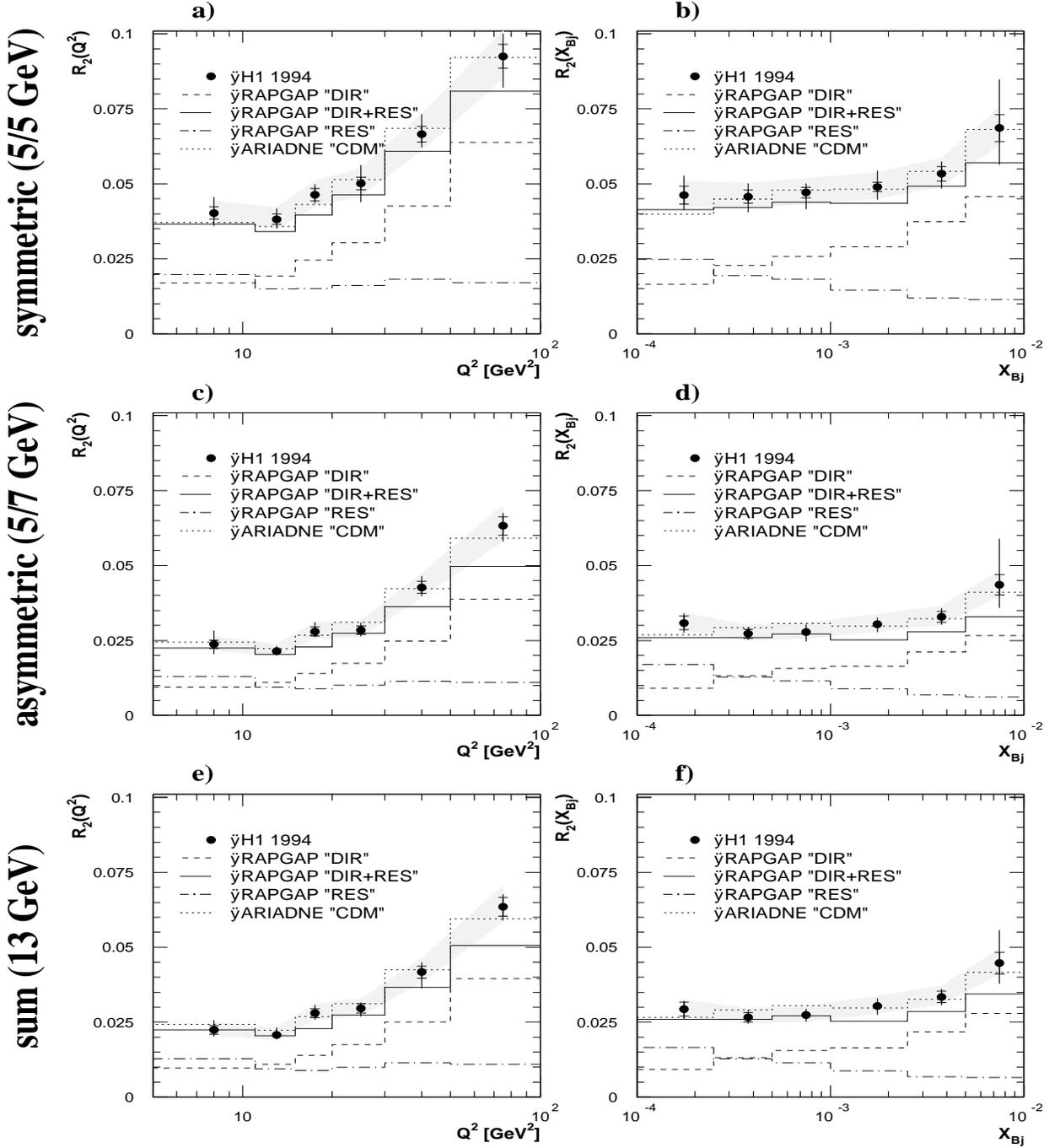,height=18cm,width=\textwidth,
              bbllx=40,bblly=70,bburx=570,bbury=800} }
\end{center}
\caption{\label{fig3}
  {\it Di-jet rate \rz\ as a function of $Q^2$ (a,c,e), integrated over
  \xbj, and as a function of $\xbj$ (b,d,f), integrated over $Q^2$
  for the symmetric (a,b), asymmetric (c,d), and sum (e,f) cut scenario 
  on the $p^*_t$ of the two jets.
  The data are corrected to the hadron level. The inner bars give the
  statistical errors, the full error bars include the bin by bin
  systematic errors. The grey band shows the overall systematic
  uncertainty due to the hadronic energy scale of the calorimeter and
  the uncertainty of the radiative correction.
  Also indicated are the predictions from MC models. 
  Their statistical errors are smaller than the statistical errors of 
  the data.} }
\end{figure}

} The di-jet
rate \rz\ is shown in Figs.~\ref{fig3}a and \ref{fig3}b as a
function of $Q^2$ and \xbj\ respectively. The data have been
corrected for detector effects to the hadron level.  The results
correspond to the phase space region defined by Eqs.
\ref{psdis}--\ref{psdeta}. The data show a jet rate rising with $Q^2$
and flat in \xbj\ except for the highest \xbj-value.

For reasons to be explained later, two further scenarios have been 
investigated, where in addition to our basic requirement of 
$p^*_t > 5$~GeV for each jet (symmetric scenario), we demand either at 
least 7~GeV for the jet with the highest $p^*_t$ (asymmetric scenario)
or at least 13~GeV for the sum of the absolute values of the jet 
transverse momenta (sum scenario). The results for \rz\ for the 
asymmetric scenario are given in Figs.~\ref{fig3}c,d and for the sum 
scenario in Figs.~\ref{fig3}e,f. Table 1 in the appendix summarizes
the di-jet rates for the three different scenarios.  

\subsection{Comparison of data with LO QCD models}

The results for the three different selections of jet phase space are 
compared to predictions from MC models based on perturbative QCD (see
previous section). The LO DIR model fails to describe 
the data as demonstrated in Figs.~\ref{fig3}a to \ref{fig3}f with 
RAPGAP\@. In particular in the region of small $Q^2$ and \xbj\ the DIR 
model underestimates the data by a factor 2--3. 
 
Choosing $\mu^2 = Q^2$ for the hard scale does not change the results
considerably.
Using different parton density parameterizations (CTEQ4L, CTEQ4A4, CTEQ4HJ, 
MRSR1, and MRSR2) leads to variations in \rz\ of up to 10\% in the lowest 
and the highest $Q^2$ and \xbj\ bins when compared to
\mbox{CTEQ4M}, our default. The world average value of
$\alpha_s(M_Z^2) = 0.118$ corresponds to  
$\Lambda^{(5)}_{\overline{MS}} = 209^{+39}_{-33}$~MeV \cite{pdg}, very close 
to the fit value of 202~MeV for CTEQ4M. With the CTEQ4M parton densities 
but with $\Lambda^{(5)}_{\overline{MS}} = 250$~MeV (an increase of about one 
standard deviation), \rz\ increases by less than 10\% in the lowest $Q^2$ 
and \xbj\ bins.
%

We conclude that a LO matrix element calculation 
assuming only direct interactions of the virtual photon
in combination with DGLAP 
parton showers as an approximation to higher order effects is not able to 
account for the observed di-jet rates.
%
 
Adding a significant contribution to the di-jet cross section
from resolving the structure of the 
virtual photon, as predicted by the DIR+RES model as implemented in RAPGAP, 
gives a good description of the data.  It should be noted, however, that 
considerable freedom exists in tuning the model to data, in particular by 
varying the choice of the hard scale, and the parton densities in the 
virtual photon.
%

The CDM model, as implemented in ARIADNE, is also able to describe the 
di-jet rate well, both in absolute value and in the $Q^2$ and \xbj\
dependence. We used a parameter setting which had been tuned to give a 
good description of transverse 
energy flows and particle spectra \cite{tuning}.
Here too, it should be remarked that by varying the 
parameters for the proton and photon size suppression (see section 2) 
within sensible limits\footnote{PARA(10) and PARA(15), default 1.0 
\cite{ariadne}; we used 1.5 and 0.5 respectively and varied them
independently between 0.5 and 1.5.}, the predictions 
of this model can be changed by up to 40\% in the lowest bin and 
about 20\% in the highest bin in $Q^2$ and \xbj. 

\subsection{Comparison of data with NLO QCD calculations}

We now investigate whether a NLO QCD calculation is able to describe the 
data.  For this purpose we have used results from the programs 
DISENT and JETVIP 
\cite{jetvip}. For the calculation of \rz\, for the 
direct or point-like coupling of the photon to the partons in 
the proton, the two programs agree to better than 5\%. As mentioned in 
section 7, JETVIP can also calculate the direct and resolved photon 
contribution in NLO.     
Both programs provide parton level cross sections rather than a full 
hadronic final state. However, the DIR and CDM models suggest that the 
hadronization effects are small for jet transverse momenta above 5~GeV.  
The di-jet rate at the parton level was found to be for LEPTO
(ARIADNE) typically 9\% (20\%) and not 
more than 12\% (25\%) higher than the rate at the hadron level.

\ifthenelse{\boolean{figatend}}{}{\begin{figure}
\begin{center}
\mbox{\epsfig{file=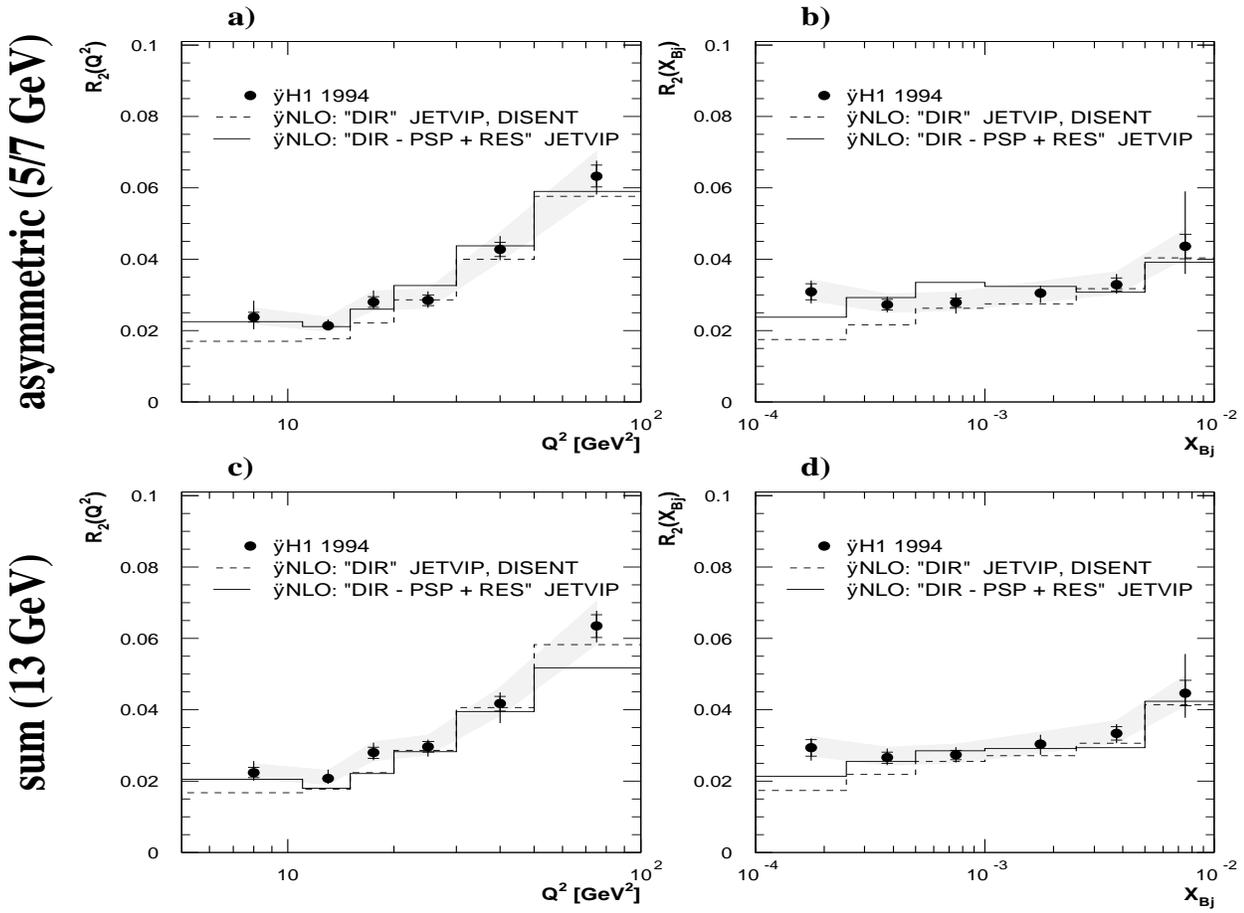,height=12.11cm,width=\textwidth,
              bbllx=40,bblly=180,bburx=570,bbury=690} }
\end{center}
\caption{\label{fig4}
  {\it Di-jet rate \rz\ as a function of $Q^2$ (a,c), integrated over
  \xbj, and as a function of \xbj\ (b,d), integrated over $Q^2$.
  The data (corrected to the hadron level) for the asymmetric (a,b) and the 
  sum scenario (c,d) are 
  compared to different NLO calculations (at the parton level). The
  data are the same as those shown in Figs.~3c to 3f.
  The statistical errors of the NLO calculations are smaller than the 
  statistical errors of the data.}  }
\end{figure}

}
Figs.~4a,b show the hadron level di-jet rate \rz\ versus $Q^2$ and 
\xbj\ for the asymmetric scenario, and in Figs.~4c,d for the sum 
scenario, compared to the NLO QCD calculation of the direct contribution
(labeled DIR in the figures) by JETVIP and DISENT\@. Good agreement 
is observed between data and the direct NLO QCD calculation, except for 
the lowest $Q^2$  and \xbj\ bin. 

The sensitivity to variations of the parton density parameterizations is 
similar to the LO DIR case discussed above.
Varying the factorization and renormalization scale $\mu^2$ by 
factors of 4 results in cross section variations of less than 20\%.
Choosing $\mu^2 = Q^2$ as the scale enhances the cross section in the lowest
\xbj\ and $Q^2$ bins by up to 30\%, improving the agreement with data.
At the same time this introduces however a large sensitivity to scale variations
(up to 50\% and 65\% in the lowest $Q^2$ and \xbj\ bins). This indicates, as
one might expect, that $Q^2$ is not the proper scale to use in a
kinematic domain where $Q^2 \ll p_t^2$. 

The agreement with data at low \xbj\ and $Q^2$ is improved when
contributions from resolving the virtual photon structure are included
in NLO (labeled DIR-PSP+RES in the figure). In order to avoid
double counting in the full NLO QCD calculation it is necessary to
subtract the contribution from the virtual photon splitting into
$q\bar{q}$, where one of the quarks subsequently interacts with a
parton from the proton to produce two high $p_t$ jets, since this
contribution is part of the parameterization of the virtual photon
structure function \cite{sasgam}. 
We refer to this perturbatively calculated contribution from
photon splitting as defined in \cite{jetvip} as PSP and the contribution from
resolving the photon structure as RES.

Two interesting observations can be made. First, at large $Q^2$ the
difference between the NLO direct part (DIR) and the full calculation
(DIR-PSP+RES) is found to be rather small, which implies that the
NLO resolved part (RES) is saturated by the contribution from virtual 
photon splitting (PSP).
Second, the full NLO calculation 
is close to the LO RAPGAP DIR+RES prediction shown as the full line
in Figs.~3c to 3f. This suggests, together with the first observation, that
the large resolved contribution needed in LO to describe the data for
the larger $Q^2$ bins is included in the NLO DIR cross section.

It should be noted that in NLO the RES contribution depends 
less on the choice of the hard scale and the parton densities in the 
virtual photon than in LO\@. This is due to the subtraction procedure
and because it is a NLO calculation \cite{jetvip}.
Of course the uncertainty due to the rather poorly known parton density
of the virtual photon remains.

A comparison of the data on \rz\ and the NLO QCD calculation for the
symmetric scenario is not shown, because the calculation for this
case is not reliable although the measurement is valid and
infrared safe. The calculations from both DISENT and JETVIP underestimate
the data and give different predictions. This can be understood as a
feature of any fixed order calculation which gives large negative
cross sections in the phase space region where both jets have almost
identical transverse momenta. The problem in
the prediction of di-jet rates for symmetric cuts on the jet
transverse momenta has been noted in the framework of the phase space
slicing method \cite{klasen-kramer} and discussed in detail for this
and the subtraction method in \cite{frixione-ridolfi}. A correct
treatment of this phase space region would need a resummation to all
orders \cite{frixione-ridolfi, catani-webber}.

\subsection{Event topology}

The conclusions of the underlying picture derived above can be checked by
a study
of the event topology. 
In the DIR+RES model the hardest emission leading to
the observed jets may occur anywhere in the ladder, as depicted in
Fig.~\ref{figlad}b.  In this case, additional hadronic activity is
expected from the virtual photon ``remnant'' and additional parton
emission from the top part of the ladder. 
\ifthenelse{\boolean{figatend}}{}{\begin{figure}
\begin{center}
\mbox{\epsfig{file=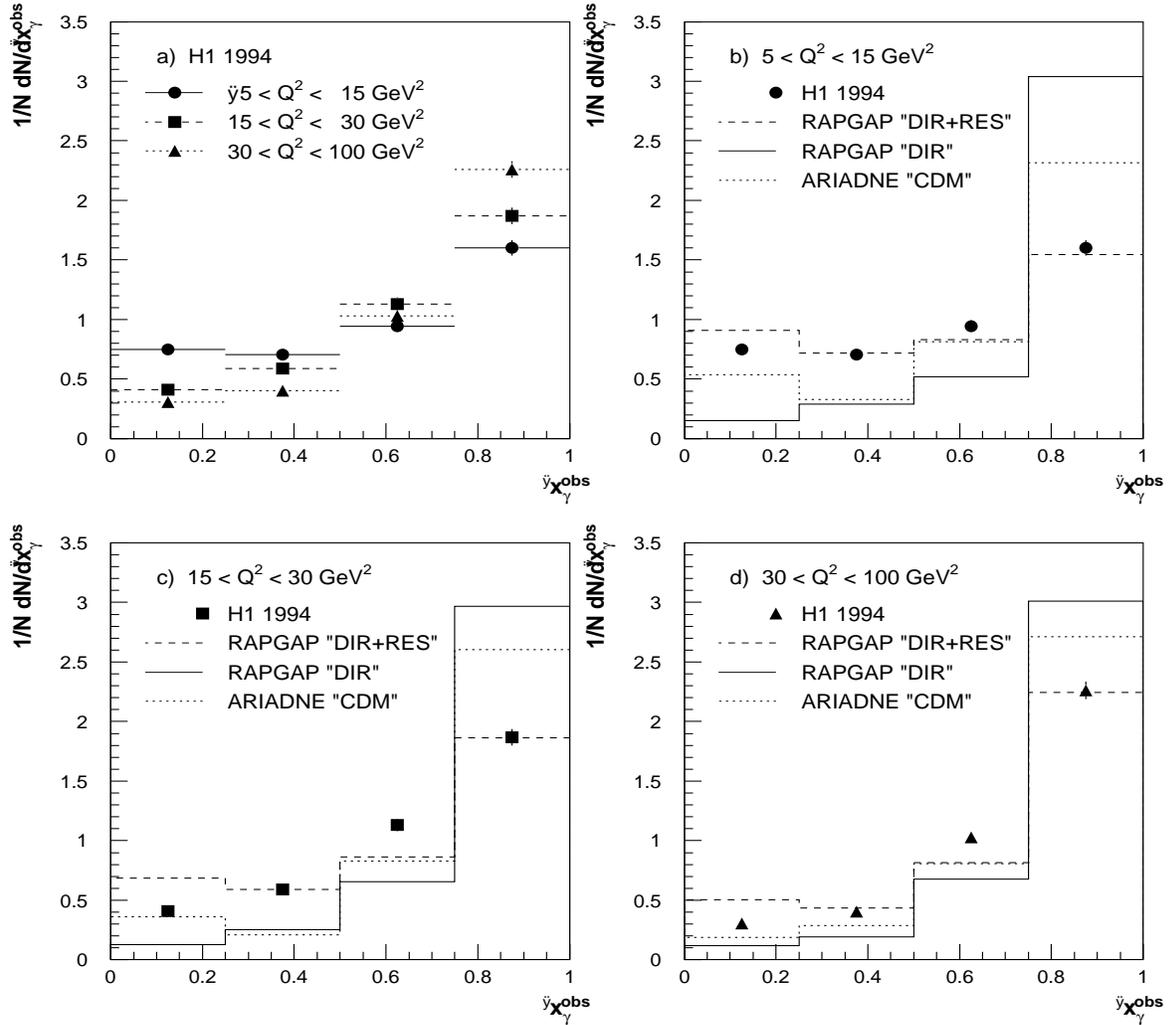,height=14cm,width=\textwidth,
              bbllx=40,bblly=180,bburx=570,bbury=690}}
\end{center}
\caption{\label{xgam}
  {\it Uncorrected distribution of $x_{\gamma}^{\rm obs}$
  in three different $Q^2$ bins (a).
  In b) to d) the data are compared for each bin in $Q^2$ to 
  the DIR model and the DIR+RES model as given by RAPGAP, and the CDM 
  model as implemented by ARIADNE\@. The figures are normalized to the number N
  of di-jet events. The error bars indicate the statistical error only. 
  } }
\end{figure}

} 
This activity is expected in the direction of the virtual photon which
corresponds to the backward region of the detector. 
Similar hadronic activity in the backward region can also be expected
from the CDM model due to the absence of $k_t$ ordering between 
the photon and the proton vertex (see section 2). Both models predict a 
$Q^2$ dependence of this effect which increases as $Q^2$ approaches zero. 

We define an observable which is
sensitive to additional energy flow in the photon direction:
\begin{eqnarray}
x_{\gamma}^{\rm obs} = \frac {\;\;\;\;\,\displaystyle\sum_{\rm jets}(E-p_z)}
{\displaystyle\sum_{\rm had.\;final\;state}\hspace*{-0.7cm}(E-p_z)}
\label{exgam}
\end{eqnarray}
In the limit $Q^2 \rightarrow 0$ and in LO this corresponds to the
fractional momentum of the parton from the photon entering the hard
subprocess and giving rise to the observed jet system.  In this picture
$1-x_{\gamma}^{\rm obs}$ corresponds to the fractional energy of the
photon remnant.

Fig.~\ref{xgam}a shows the uncorrected distribution
of $x_{\gamma}^{\rm obs}$ for data in three different ranges of $Q^2$.
The MC events which were used for comparison in Fig.~\ref{xgam} have
been subject to a detailed simulation of the H1 detector. In the data
an increase at low $x_{\gamma}^{\rm obs}$ is noticed as $Q^2$ decreases.  
No such effect is seen for the DIR model as represented by RAPGAP 
(full line in Figs.~\ref{xgam}b to \ref{xgam}d). 
The DIR+RES model of RAPGAP is able to give a reasonable description of both
shape and $Q^2$ dependence (dashed line in Figs.~\ref{xgam}b to \ref{xgam}d). 
The CDM model shows a similar $Q^2$
dependence but fails to describe the shape of the distribution
(dotted line in Figs.~\ref{xgam}b to \ref{xgam}d). 

\section{Conclusions}

Di-jet event rates have been measured in deep-inelastic scattering at small
\xbj\ ($ 10^{-4}$ \raisebox{-.8ex} {$\stackrel{\textstyle<}{\sim}$}
\xbj\ \raisebox{-.8ex} {$\stackrel{\textstyle<}{\sim}$} $10^{-2} $)
and moderate $Q^2$ (5~\raisebox{-.8ex}
{$\stackrel{\textstyle<}{\sim}$} $Q^2$ \raisebox{-.8ex}
{$\stackrel{\textstyle<}{\sim}$} $100 \mbox{ GeV}^2 $).
Three 
different scenarios of cuts on the transverse momenta of the jets
have been investigated: the basic symmetric requirement ($p_{t,\rm
  min}^{*}\geq 5{\rm\,GeV}$ for both jets), and additionally the
asymmetric ($p_{t,\rm
  min}^{*}\geq 5$ and 7\,GeV) and sum $p_t^*$ ($\geq 13{\rm\,GeV}$) requirements.
The analysis was performed in the hadronic centre of mass frame.
The data have been corrected for detector and QED radiative effects.
This analysis probes a region in jet phase space of small \xbj\
and jet transverse momenta squared of similar size or larger than the 
photon virtuality ($p_t^2/Q^2$ \raisebox{-.8ex}
{$\stackrel{\textstyle>}{\sim}$}1). 

Assuming a direct or point-like 
photon, leading order matrix element calculations in combination with
parton showers as an approximation of higher order effects
fail completely to describe the data. 

Adding to the leading order model additional contributions
from resolving the partons inside the virtual photon (RAPGAP) appear to give
an effective description of higher order effects leading to good
agreement with the data for all three scenarios. 
A similarly good agreement with the di-jet event rates is observed for
the colour dipole model (ARIADNE) with its features of gluon emission.
  
Next-to-leading order calculations in $\alpha_s$ 
assuming a point-like virtual photon provide a 
good description of the data for the scenarios with the asymmetric
and the sum $p^*_t$ cut, except for the lowest bin in $Q^2$ and \xbj.
This is improved by a NLO calculation which also considers contributions 
from resolving virtual photon structure.



\vspace*{-0.5\baselineskip}
\section*{Acknowledgements}

We are grateful to the HERA machine group whose outstanding efforts
have made and continue to make this experiment possible.  We thank the
engineers and technicians for their work in constructing and now
maintaining the H1 detector, our funding agencies for financial
support, the DESY technical staff for continual assistance, and the
DESY directorate for the hospitality which they extend to the non-DESY
members of the collaboration. We gratefully acknowledge fruitful
discussions with S.~Frixione, G.~Kramer, L.~L\"onnblad, B.~P\"otter, 
and J.~Rathsman.


\vspace*{-0.5\baselineskip}
\section*{Appendix}
\begin{center}
{\footnotesize \renewcommand{\arraystretch}{1.1}
\begin{tabular}{|r@{\hspace*{2.5ex}--\hspace*{1ex}}r|l|l|l|l|l|l|l|l|l|}\hline
  \multicolumn{2}{|c|}{}
& \multicolumn{3}{|c|}{symmetric (5/5~GeV)}
& \multicolumn{3}{|c|}{asymmetric (5/7~GeV)}
& \multicolumn{3}{|c|}{sum (13~GeV)} \\ \hline 
\multicolumn{2}{|c|}{$Q^2$ [GeV$^2$]}       
&  \multicolumn{1}{c|}{R$_2$}  &\multicolumn{1}{c|}{$\sigma_{\rm stat}$}&\multicolumn{1}{c|}{$\sigma_{\rm syst}$}
&  \multicolumn{1}{c|}{R$_2$}  &\multicolumn{1}{c|}{$\sigma_{\rm stat}$}&\multicolumn{1}{c|}{$\sigma_{\rm syst}$}
&  \multicolumn{1}{c|}{R$_2$}  &\multicolumn{1}{c|}{$\sigma_{\rm stat}$}&\multicolumn{1}{c|}{$\sigma_{\rm syst}$}
\\\hline
5 & 11  
&0.040 &$\pm0.002$&$_{ -0.004}^{ +0.005}$
&0.024 &$\pm0.001$&$_{ -0.003}^{ +0.004}$
&0.022 &$\pm0.001$&$_{ -0.002}^{ +0.003}$
\\\hline
11& 15  
&0.038 &$\pm0.002$&$_{ -0.002}^{ +0.003}$
&0.021 &$\pm0.001$&$_{ -0.001}^{ +0.001}$
&0.021 &$\pm0.001$&$_{ -0.001}^{ +0.002}$
\\\hline
15& 20  
&0.046 &$\pm0.002$&$_{ -0.002}^{ +0.003}$
&0.028 &$\pm0.002$&$_{ -0.001}^{ +0.003}$
&0.028 &$\pm0.002$&$_{ -0.002}^{ +0.002}$
\\\hline
20& 30  
&0.050 &$\pm0.002$&$_{ -0.006}^{ +0.006}$
&0.028 &$\pm0.001$&$_{ -0.002}^{ +0.002}$
&0.030 &$\pm0.002$&$_{ -0.002}^{ +0.001}$
\\\hline
30& 50  
&0.067 &$\pm0.003$&$_{ -0.004}^{ +0.006}$
&0.043 &$\pm0.002$&$_{ -0.002}^{ +0.003}$
&0.042 &$\pm0.002$&$_{ -0.005}^{ +0.003}$
\\\hline
50& 100 
&0.093 &$\pm0.004$&$_{ -0.010}^{ +0.006}$
&0.063 &$\pm0.003$&$_{ -0.004}^{ +0.003}$
&0.063 &$\pm0.003$&$_{ -0.003}^{ +0.003}$
\\\hline
\end{tabular}\hspace{2ex}
\vspace*{0.5cm}

\begin{tabular}{|r@{$\;$--$\;$}r|l|l|l|l|l|l|l|l|l|}
\hline
  \multicolumn{2}{|c|}{}
& \multicolumn{3}{|c|}{symmetric (5/5~GeV)}
& \multicolumn{3}{|c|}{asymmetric (5/7~GeV)}
& \multicolumn{3}{|c|}{sum (13~GeV)} \\ \hline 
\multicolumn{2}{|c|}{\xbj}       
&  \multicolumn{1}{c|}{R$_2$}  &\multicolumn{1}{c|}{$\sigma_{\rm stat}$}&\multicolumn{1}{c|}{$\sigma_{\rm syst}$}
&  \multicolumn{1}{c|}{R$_2$}  &\multicolumn{1}{c|}{$\sigma_{\rm stat}$}&\multicolumn{1}{c|}{$\sigma_{\rm syst}$}
&  \multicolumn{1}{c|}{R$_2$}  &\multicolumn{1}{c|}{$\sigma_{\rm stat}$}&\multicolumn{1}{c|}{$\sigma_{\rm syst}$}
\\\hline
        $10^{-4}$ & $2.5\cdot10^{-4}$ 
& 0.046 & $\pm0.003$  & $_{ -0.004}^{ +0.006}$
& 0.031 & $\pm0.002$  & $_{ -0.002}^{ +0.002}$
& 0.029 & $\pm0.002$  & $_{ -0.003}^{ +0.001}$
\\\hline
$2.5\cdot10^{-4}$ & $5.0\cdot10^{-4}$ 
& 0.046 & $\pm0.002$  & $_{ -0.005}^{ +0.004}$
& 0.027 & $\pm0.002$  & $_{ -0.001}^{ +0.002}$
& 0.027 & $\pm0.002$  & $_{ -0.001}^{ +0.002}$
\\\hline
$5.0\cdot10^{-4}$ & $10^{-3}$         
& 0.047 & $\pm0.002$  & $_{ -0.005}^{ +0.003}$
& 0.028 & $\pm0.001$  & $_{ -0.003}^{ +0.002}$
& 0.027 & $\pm0.001$  & $_{ -0.002}^{ +0.002}$
\\\hline
        $10^{-3}$ & $2.5\cdot10^{-3}$ 
& 0.049 & $\pm0.002$  & $_{ -0.004}^{ +0.005}$
& 0.030 & $\pm0.001$  & $_{ -0.002}^{ +0.002}$
& 0.030 & $\pm0.001$  & $_{ -0.003}^{ +0.002}$
\\\hline
$2.5\cdot10^{-3}$ & $5.0\cdot10^{-3}$ 
& 0.053 & $\pm0.002$  & $_{ -0.004}^{ +0.003}$
& 0.033 & $\pm0.002$  & $_{ -0.002}^{ +0.002}$
& 0.033 & $\pm0.002$  & $_{ -0.002}^{ +0.002}$
\\\hline
$5.0\cdot10^{-3}$ & $10^{-2}$         
& 0.069 & $\pm0.005$  & $_{ -0.011}^{ +0.016}$
& 0.044 & $\pm0.003$  & $_{ -0.007}^{ +0.015}$
& 0.045 & $\pm0.004$  & $_{ -0.006}^{ +0.010}$
\\\hline
\end{tabular}
}
\end{center}
Table 1: {\it Di-jet rate in bins of $Q^2$ and \xbj, and statistical and
  systematic errors for the symmetric, the asymmetric, and the sum 
  cut scenario on the $p^*_t$ of the two jets. 
  The overall systematic  error of
  $_{-8\%}^{+10\%}$ for the symmetric scenario and
  $_{-8\%}^{+11\%}$ for the asymmetric and sum scenario,
  arising from the uncertainty of the hadronic
  energy scale of the calorimeter and the uncertainty of the radiative QED 
  corrections, is not included.}
\label{tab_r2}


\end{document}